\documentclass[aps,twocolumn,prl,amsmath,amssymb,floatfix,superscriptaddress]{revtex4-1}
\usepackage{graphicx}
\usepackage{bm}
\usepackage{color}
\usepackage{hyperref}
\hypersetup{
  colorlinks   = true, 
  urlcolor     = blue, 
  linkcolor    = blue, 
  citecolor   = red 
}
\usepackage{epsfig}
\usepackage{epstopdf}

\begin{document}

\title{High-performance frequency stabilization of ultraviolet diode lasers by using dichroic atomic vapor spectroscopy and transfer cavity}
\author{Danna Shen}
\author{Liangyu Ding}
\author{Qiuxin Zhang}
\author{Chenhao Zhu}
\author{Yuxin Wang}
\affiliation{Department of Physics, Renmin University of China, Beijing 100872, China}
\author{Wei Zhang}
\author{Xiang Zhang}
\thanks{Corresponding author. E-mail:~siang.zhang@ruc.edu.cn}
\affiliation{Department of Physics, Renmin University of China, Beijing 100872, China}
\affiliation{Beijing Key Laboratory of Opto-electronic Functional Materials and Micro-nano Devices,Renmin University of China, Beijing 100872, China}

\begin{abstract}
Ultraviolet (UV) diode lasers are widely used in many photonics applications. But their frequency stabilization schemes are not as mature as frequency-doubling lasers, mainly due to some limitations in the UV spectral region. Here we developed a high-performance UV frequency stabilization technique implemented directly on UV diode lasers by combining the dichroic atomic vapor laser lock and the resonant transfer cavity lock. As an example, we demonstrate a stable locking with frequency standard deviations of approximately 200 KHz and 300 KHz for 399nm and 370nm diode lasers in 20 minutes. We achieve a long-term frequency drift of no more than 1 MHz for the target 370nm laser within an hour, which was further verified with fluorescence counts rates of a single trapped $^{171}$Yb$^+$ ion. We also find strong linear correlations between lock points and environmental factors such as temperature and atmospheric pressure.
\end{abstract}

\maketitle

\section{Introduction}

Ultraviolet (UV) lasers are increasingly used in numerous photonics applications in research and industry, including quantum information processing, atomic frequency standards, photoemission spectroscopy, and laser photolithography. Some applications have very stringent requirements for the linewidth and the center frequency of laser, e.g., the experiments related to cold atoms or trapped ions\cite{heinzen_quantum-limited_1990}. Recently, several conventional UV laser diodes and their counterpart external cavity grating feedback diode lasers (ECDLs) have become widely available. They are more compact and lightweight, more energy-efficient, require less maintenance, and cost much lower than traditional frequency-doubling UV lasers. However, since UV-coated optics are relatively inefficient and easily damaged, the frequency stabilization for UV diode lasers is usually more difficult than traditional UV lasers, for which the frequency stabilization can actually be done in the visible or near-infrared (NIR) spectral region by using the excess longer wavelength laser.

In general, a laser frequency stabilization system should at least consist of an absolute frequency reference and a frequency locking component. Commonly used absolute frequency references include absorption lines, other stable lasers and Fabry-Perot (FP) cavities, especially high-finesse cavities made of ultra-low thermal expansion (ULE) materials and installed in specially designed stable vacuum housings. Corresponding frequency locking schemes include the saturation absorption dither locking (SADL)\cite{zhou_frequency-stabilized_2010,zhang_compact_2012,dammalapati_saturated_2009,zhang_novel_2009}, the polarization spectroscopy (PS)\cite{zhu_polarization_2014,lee_frequency_2014}, the modulation transfer spectroscopy (MTS)\cite{ru_flexible_2018,wu_modulation_2018,cheng_laser_2014,wang_frequency_2011,mccarron_modulation_2008,noh_modulation_2011}, the dichroic atomic vapor laser lock (DAVLL)\cite{kim_frequency-stabilized_2003,corwin_frequency-stabilized_1998}, the optical injection lock (OIL), the optical phase locked loop lock (OPLL), the Pound-Drever-Hall (PDH) technique\cite{wang_artificial_2019}, and the scanning transfer cavity lock\cite{uetake_frequency_2009,subhankar_microcontroller_2019,jackson_laser_2018,bohlouli-zanjani_optical_2006}, etc. However, the widely used rich absorption lines of Rb or iodine\cite{eloranta_frequency_2006}, as well as the frequency-shifting fiber electro-optic modulator (EOM) with a wide tuning range, are generally not available in the UV spectral region. It is also usually difficult to coat a transfer FP cavity in the operating wavelength range from UV to NIR. Therefore, for UV diode lasers with relatively limited output power, the best practice should be a UV absorption line scheme, but this only applies if the target lock point is fortunately located near a strong absorption line.

In order to overcome these limitations, we developed a high-performance frequency stabilization technique that combines the DAVLL and the resonant transfer cavity lock. The new approach needs an auxiliary UV laser with less strict center frequency requirement. We first lock the auxiliary laser to a strong absorption spectral line using the DAVLL method, then the frequency lock is passed to the target UV laser through the resonant transfer cavity. The lock point of the target UV laser can be fine tuned by adjusting the lock point of the DAVLL in a huge real-time frequency tuning range up to GHz, which can cover the whole free spectral range (FSR) of conventional FP cavities. Thus we can lock the target UV laser to narrow linewidth at any center frequency. All frequency locking and shifting functions, including automatic lock-point finding and automatic two-way top-of-fringe transfer lock, are implemented by digital signal processing algorithms\cite{feng_arbitrary_2015,jorgensen_simple_2016} based on field-programmable gate array (FPGA) technology. We build and test this approach in a trapped ion experiment. The main apparatus is an Yb hollow-cathode lamp (HCL) for the DAVLL and a conventional FP cavity for the resonant transfer cavity lock. By analyzing and cancelling laser power and polarization jittering, we achieve frequency standard deviations of $202$~KHz and $309$~KHz for the auxiliary 399nm laser and the target 370nm laser, respectively. Furthermore, we analyze the relationship between lock points and environmental parameters, e.g., temperature and atmospheric pressure, and find strong linear correlations that can be reasonably explained. Finally, we achieve a long-term frequency drift of no more than 1 MHz for the 370nm laser and verify it by recording and analyzing fluorescence counts rate of a single trapped $^{171}$Yb$^+$ ion.

\section{Experimental setup}

\begin{figure*}[htbp]
\centering
\includegraphics[width=\linewidth]{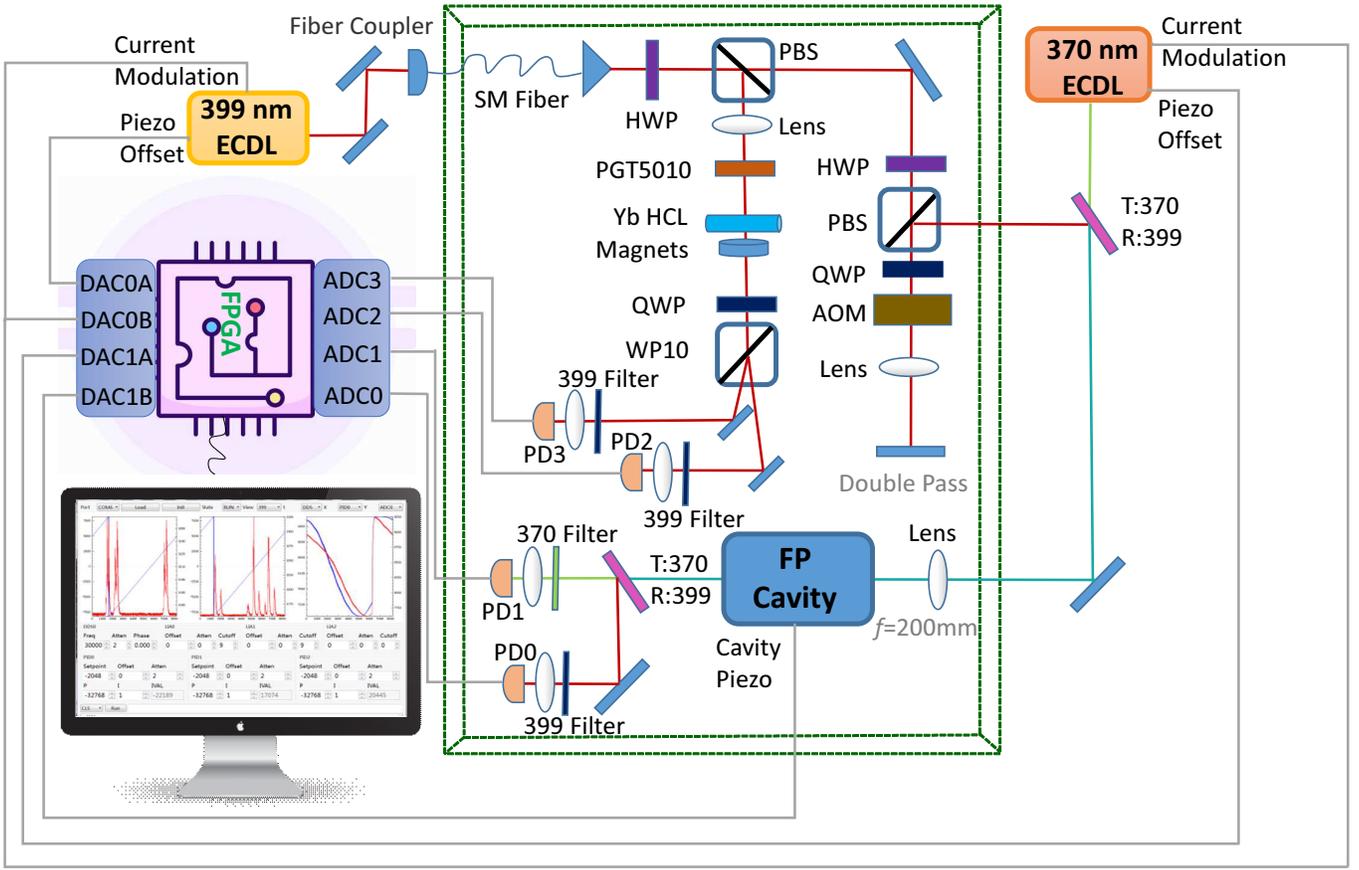}
\caption{(Color online) Experimental setup. AOM, acousto-optic modulator; ECDL, external cavity dioder laser; HCL, hollow-cathode lamp; HWP; half-wave plate; PBS, polarization splitting prism; PD, photodetector; PGT5010, Glan-Taylor prism; QWP, quarter-wave plate; WP10, Wollaston prism.}
\label{fig1}
\end{figure*}

The entire experimental system is placed in a closed box as schematically shown in Fig.~\ref{fig1}.
The 399nm laser beam enters the box through a polarization-maintaining fiber (SM-300), and is then divided into two beams by a half-wave plate (HWP) and a polarized-beam splitter (PBS), which are used for the DAVLL and the transfer cavity, respectively.
The 399nm beam for the DAVLL passes a Glan-Taylor prism (PGT5010) with a high extinction ratio to ensure that the output beam is purely linear polarized, and then hits the center of the Yb HCL.
A pair of permanent magnet rings are placed close to the HCL to generate a stable magnetic field, which causes a relatively large Zeeman splitting for the atoms in the HCL.
Left-handed and right-handed circularly polarized beams pass through a quarter-wave plate (QWP) and become vertically or horizontally polarized.
The two polarized beams are then separated by a Wollaston prism (WP10), and measured by photodetectors (DET100).
We install narrow-linewidth optical filters and short-focus lens with adjustable distance in front of each photodetector, which significantly improves the signal-to-noise ratio of PD signals by reducing the interference of background light and increasing the beam intensity shining on the PD.
The other 399nm beam for the transfer cavity is combined with a weak ($<10~\mu$W) 370nm laser beam through a dichroic mirror. The double pass AOM path is used for frequency shifting of testing purpose and can be optional.
The combined beam enters the confocal FP cavity (model SA200-3B) through a lens for pattern matching. The transmitted 370nm and 399nm beams are separated again by a dichroic mirror and measured with PDs.

The analog voltage signals of PDs are converted to digital signals through an eight-channel 16-bit analog-to-digital converter (ADC, AD7606). These signals undergo a series of digital signal processing, then convert into feedback voltages by two two-channel 16-bit digital-to-analog converters (DAC, DAC8563), and finally output to the piezoelectric ceramics of the laser or the FP cavity to implement the stabilization of frequency or cavity length. Digital signal processing algorithms including digital signal synthesizers, digital lock-in amplifiers, digital low-pass filters, and digital proportional-–integral-–derivative (PID) controllers, are implemented through FPGA (Spartan 6 XC6SLX9, $100$~MHz clock) programming on a low-cost Mojo V3 development board. The all-digital design greatly reduces the impact of noise during signal processing and improves robustness. All the hardware modules, including the transfer function generator for testing purpose, are wrapped into standard trigger-driven input/output interfaces. And all the logical connections can be reconfigured on the fly from our PC program, which can also communicate with the FPGA and automatically find the lock point. We only need to adjust some parameters on the computer via a user-friendly interface and confirm the position of the lock point to complete the entire frequency stabilization process. The overall feedback bandwidth is around $100$~KHz mainly limited by the maximum sampling speed of the ADC.

\section{Dichroic atomic vapor spectroscopy}

The DAVLL method subtracts the absorption signals of Zeeman level transitions as the frequency discriminator error signal, and consumes much less laser power than other pump-based spectroscopy methods. We choose a low-cost HCL (model Yb HL-1) with an operating voltage of approximately $132$~V and an operating current of approximately $5$~mA. We use its absorption line near $398.9$~nm as the absolute frequency reference. Using HCL instead of evacuated heat pipe or vacuum chamber makes the apparatus much simpler, and potentially enables self-test for the target 370nm laser since the HCL has both atoms and ions inside.

\begin{figure*}[htbp]
\centering
\includegraphics[width=0.7\linewidth]{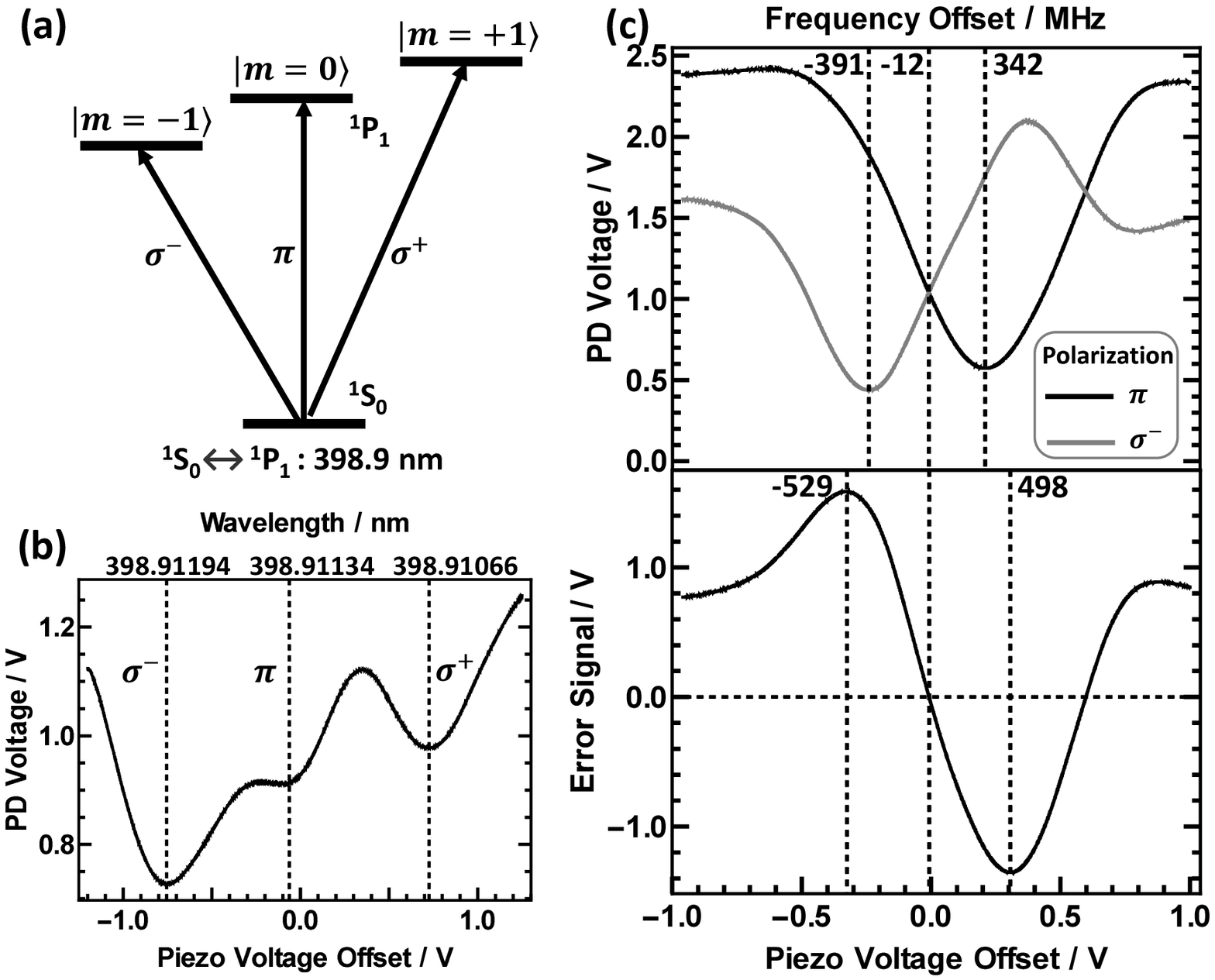}
\caption{(a) The level structure of Zeeman levels within the excited manifold $^1P_1$ of the Yb atom; (b) The Zeeman level absorption valleys (black dashed) corresponding to the three transitions shown in (a) can be well identified in the absorption spectrum; (c) The absorption spectrum of the $\pi$ (top, black solid) and $\sigma^-$ (top, gray solid) polarization, and the error signal (bottom, black solid).}
\label{fig2}
\end{figure*}

As shown in Fig.~\ref{fig2}~(a), the excited state $^1P_1$ of the Yb atom contains three degenerate Zeeman energy levels with energies 
\begin{eqnarray}
\Delta E(^1P_1)/\hbar=(14.5~\mathrm{MHz}/\mathrm{mT})mB  \quad ({\rm for\ }m=\pm 1).
\end{eqnarray}
We first measure the absorption spectrum obtained from PD2 by scanning the piezo voltage offset of the 399nm laser. For magnets at different positions, the wavelength of the central $m=0$ absorption valley is always stable at $398.91134$~nm. We adjust the position of magnets and wave plates until all three absorption valleys of Zeeman transitions can be identified clearly, as shown in Fig.~\ref{fig2}~(b). The magnetic field intensity produced by permanent magnets at the center of the HCL is approximately $76$~mT (measured without HCL). This corresponds to a frequency gap of approximately $1.1$~GHz, which is in consistence with the measured wavelength difference between $m=0$ and $m=-1$ transitions. The $200$~MHz additional frequency gap between $m=1$ and $m=0$ transitions is believed to be caused by isotope shifts. However, we notice that it is difficult to observe clear absorption spectrum of both circular polarizations simultaneously, which may also explain an abnormal half frequency splitting as mentioned in Ref.~\cite{kim_frequency-stabilized_2003}. Instead, we use absorption signals of $m=0$ and $m=-1$, and scan the piezo voltage offset of the 399nm laser at the center wavelength $398.91154$~nm. The top panel of Fig.~\ref{fig2}~(c) shows the signal of $V_\mathrm{ADC2}$ and $V_\mathrm{ADC3}$, which are respectively absorption signals of the $\pi$ and $\sigma ^ -$ polarization beams obtained from PD2 and PD3. The error signal $V_\mathrm{ADC23}$ is obtained by directly subtracting the two signals, as shown in the bottom panel of Fig.~\ref{fig2}~(c). The slope of this error signal is $265$~MHz/V, which is obtained by linear fitting the center portion. Thus with our $\pm 10$~V 16-bit ADC, the minimum detectable laser frequency change in the DAVLL setup is $81$~KHz.

However, this original error signal of the DAVLL method may be affected by beam power, polarization and optical path jitter. Therefore, we use polarization-maintaining fibers instead of free space to conduct the laser beam to the HCL in the closed box, and use $(V_\mathrm{ADC2}-V_\mathrm{ADC3}) / (V_\mathrm{ADC2}+V_\mathrm{ADC3}-V_0)$ instead of $V_\mathrm{ADC2}-V_\mathrm{ADC3}$ as the error signal to eliminate the jitter of optical power, where $V_0$ is the sum of the PDs' dark voltages. This normalized error signal is processed by the PID controller and converted to voltage by the DAC, and then fed back to the piezoelectric ceramics of the 399nm laser to compensate fluctuations in laser frequency.

\begin{figure*}[htbp]
\centering
\includegraphics[width=\linewidth]{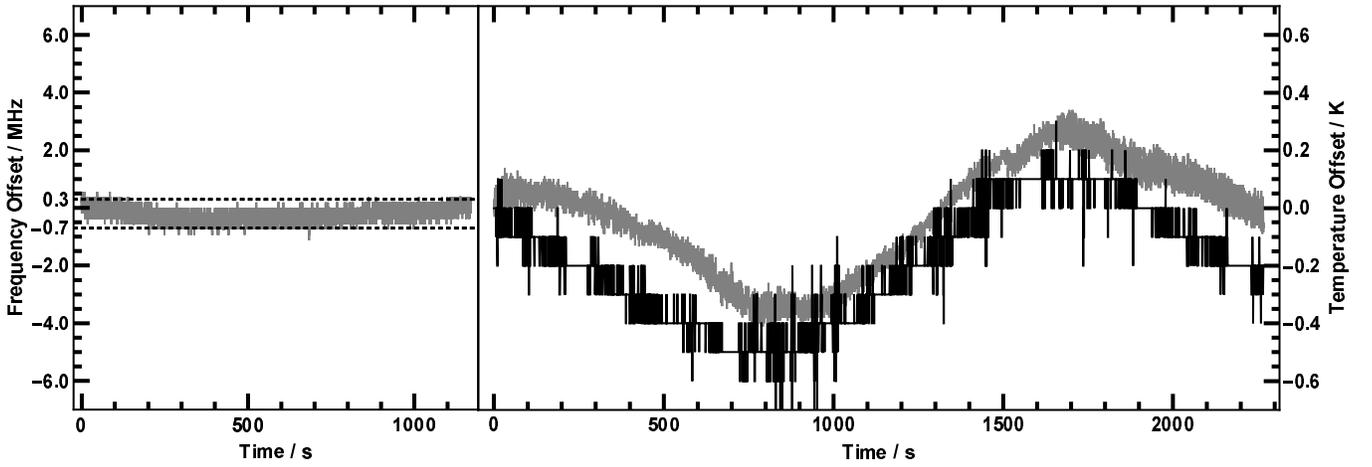}
\caption{Relative frequency offsets of the 399nm laser (gray) for (left) a short-term within 20 minutes and (right) a long-term within 40 minutes measurement as well as relative temperature offsets in the box (black).}
\label{fig3}
\end{figure*}

The performance of our frequency stabilization scheme for the 399nm laser is measured with a sub-MHz resolution wavelength meter (model WS8), as shown in the left panel of Fig.~\ref{fig3}. The frequency jitter is limited within $1$~MHz ($-0.7\sim 0.3$~MHz) in 20 minutes, and the standard deviation is $202$~KHz, which is close to the instantaneous linewidth of the diode laser. Over a longer period of time, a linear correlation between frequency drift and temperature change inside the box is observed, as shown in the right panel of Fig.~\ref{fig3}. The temperature coefficient of the lock point change obtained by linear fitting is $8.458$~MHz/K, which indicates a twice coefficient $16.9$~MHz/K for the $\sigma^-$ polarization absorbtion or $1.16$~mT/K for the magnetic field intensity in the HCL. Based on this observation, we install a temperature heating controller in the box and keep the temperature around $302$~K with $0.1$~K precision. Since the relative temperature coefficient of permanent magnets' remanence is negative and is at most $-0.05$~\%/K, we believe that the dominant reason for this $1.6$~\%/K temperature coefficient should be the distance between the permanent magnets and the HCL mainly due to thermal expansion and contraction. 

\section{Transfer cavity}

\begin{figure*}[htbp]
\centering
\includegraphics[width=0.7\linewidth]{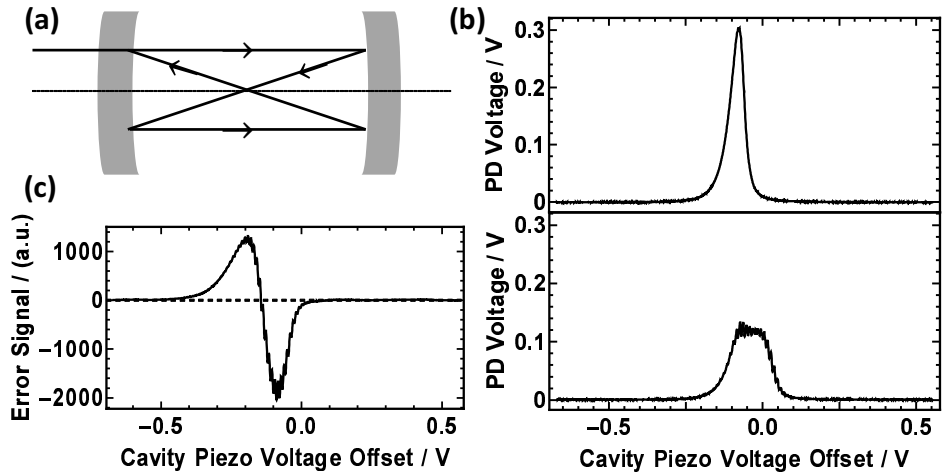}
\caption{(a) Diagram of the confocal FP cavity; (b) The transmitted peak signal without (top) and with (bottom) modulating the phase of the laser when scanning the cavity length; (c) The demodulated error signal when scanning the cavity length.}
\label{fig4}
\end{figure*}

The resonant transfer cavity lock scheme is implemented by first locking the FP cavity length to the stabilized 399nm laser and then locking the 370nm laser to the FP cavity length. For both cases we use the simple top-of-fringe method instead of the PDH method and obtain the derivative of the transmitted peak signal by modulating the phase of the laser. The slope-shape demodulated differential signal holds an approximate linear dependence on the incident laser frequency or the cavity length over the entire peak range, hence can be used as the error signal, as shown in Fig.~\ref{fig4}. By locking at the zero point of derivative, the resonant transfer cavity lock effectively avoids possible non-linearity and noise of timing or electronic levels, and has a much higher sensitivity and feedback bandwidth than the common used scanning transfer cavity lock.

After the frequency stabilization of the 399nm laser is achieved, we first scan the length of the FP cavity to a frequency scanning range equal to the FSR of $1.5$~GHz, and find a single transmitted peak signal.
A current modulation of $5$~KHz and $0.05$~mA is applied to the 399nm laser, and the demodulated slope-shape error signal is fed back to the length of the FP cavity through a PID controller.
After the cavity length is stabilized, we scan the piezo voltage offset of the 370nm laser to find its transmitted peak that resonates with the cavity length. The same current modulation is also applied to the 370nm laser and the demodulated slope-shape error signal is fed back to the piezo voltage offset of the 370nm laser through another PID controller. By adjusting the lock point wavelength of the 399nm laser to $398.911691$~nm, the 370nm laser is transfer locked at $369.526275$~nm, which is the optimal Doppler cooling frequency for our trapped $^{171}$Yb$^+$ ion setup.

\begin{figure*}[htbp]
\centering
\includegraphics[width=\linewidth]{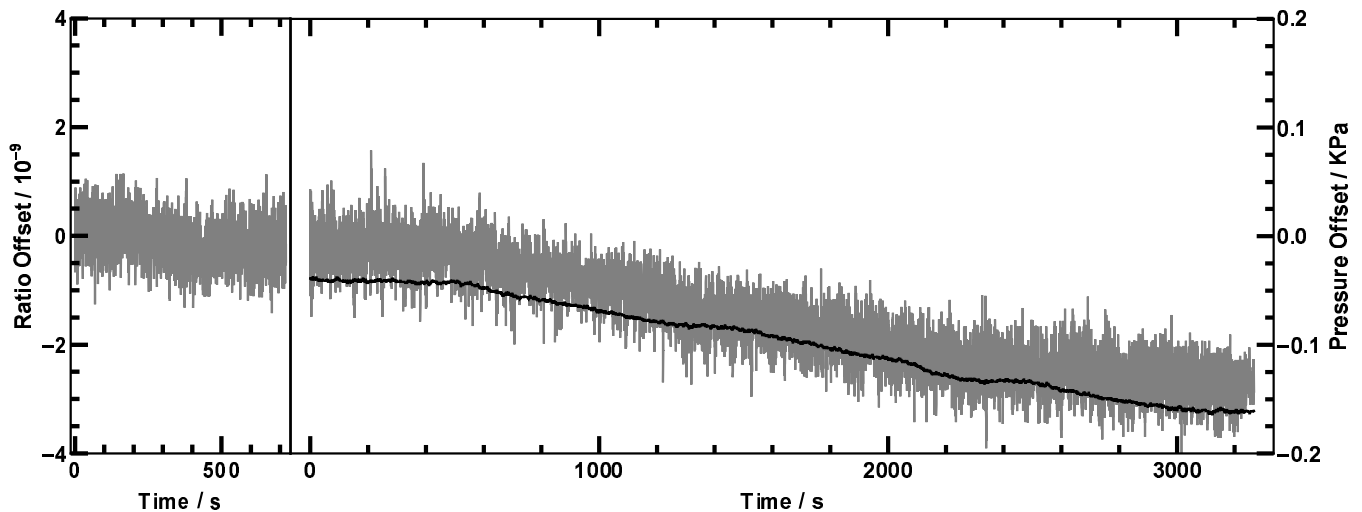}
\caption{Relative offsets of frequency ratio between the 399nm and 370nm lasers (gray) during (left) a short-term and (right) a long-term measurement as well as relative offsets of the atmospheric pressure in the box (black).}
\label{fig5}
\end{figure*}

The performance of the transfer cavity is characterized by monitoring frequency ratio between the 399nm and 370nm lasers. The frequency ratio is very stable for a time period less than 10 minutes, as shown in the left panel of Fig.~\ref{fig5}, indicating that the transfer relationship is trustworthy. Over a longer period of time, a linear correlation between the drift of the frequency ratio and the atmospheric pressure inside the box is observed, which is caused by different refractive indices changes between the 399nm and 370nm lasers, as shown in the right panel of Fig.~\ref{fig5}. The linear coefficient is fitted to be $2.05\times 10^{-8}$~/KPa, or $-2.39\times 10^{-8}$/KPa for reverse frequency ratio, which corresponds to $-18$~MHz/KPa frequency drift for the 370nm laser assuming the 399nm laser is stable. To minimize induced fluctuations, we try to keep both the laboratory door and the box airtight, and obtain a relatively stable atmospheric pressure between $101.76\sim 101.79$~KPa.

\begin{figure*}[htbp]
\centering
\includegraphics[width=\linewidth]{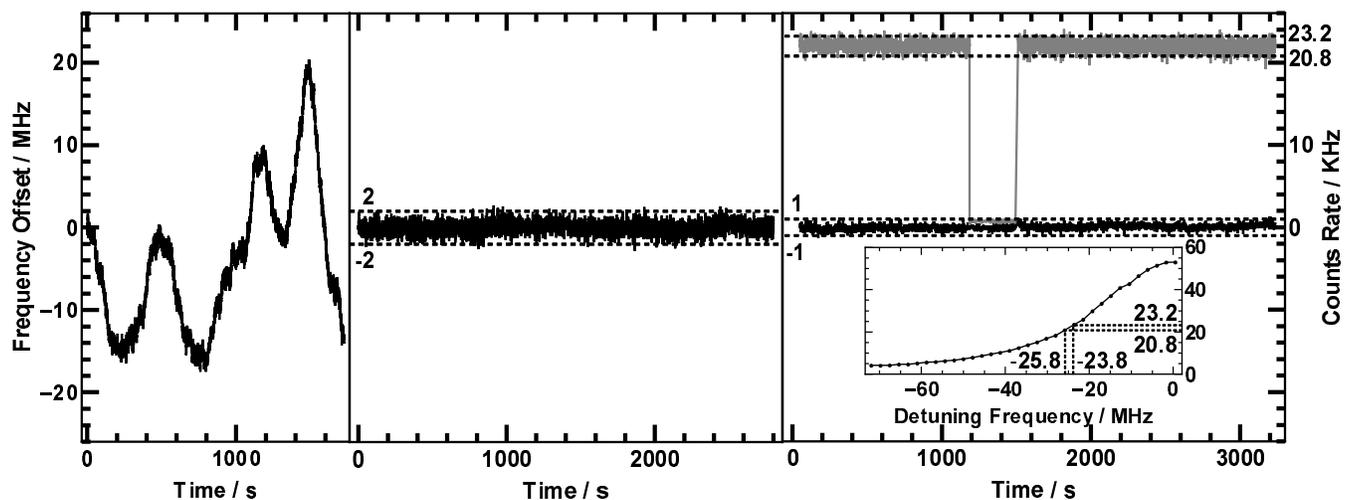}
\caption{The left, middle and right panels show the frequency drifting (black) when the laser is not locked, locked with a wavelength meter, and locked using our method, respectively. The right panel shows a stable ion fluorescence counts rate (gray) and the inset shows the fluorescence spectrum with respect to detunning frequency in MHz. The "jump" of ion fluorescence counts rate indicates an occasional ion "dark" event.}
\label{fig6}
\end{figure*}

With stabilized temperature and atmospheric pressure, we can finally stabilize the target 370~nm laser in a relatively long time scale. In Fig.~\ref{fig6}, we show the frequency drifting of the 370~nm laser under different conditions: laser is not locked, locked with a wavelength meter, and locked using our method. The standard deviations are $8.768$~MHz, $685$~KHz and $309$~KHz, respectively. The result of our method is confirmed by monitoring fluorescence counts rates of a single trapped $^{171}$Yb$^+$ ion, as shown in the right panel. Note that most fluorescence counts rates are between $20.8\sim 23.2$~KHz, so the long-term frequency drift of the 370nm laser must be in the $\pm 1$~MHz range ($-25.8\sim -23.8$~MHz) according to the measured fluorescence spectrum.

\begin{figure}[htbp]
\centering
\includegraphics[width=\linewidth]{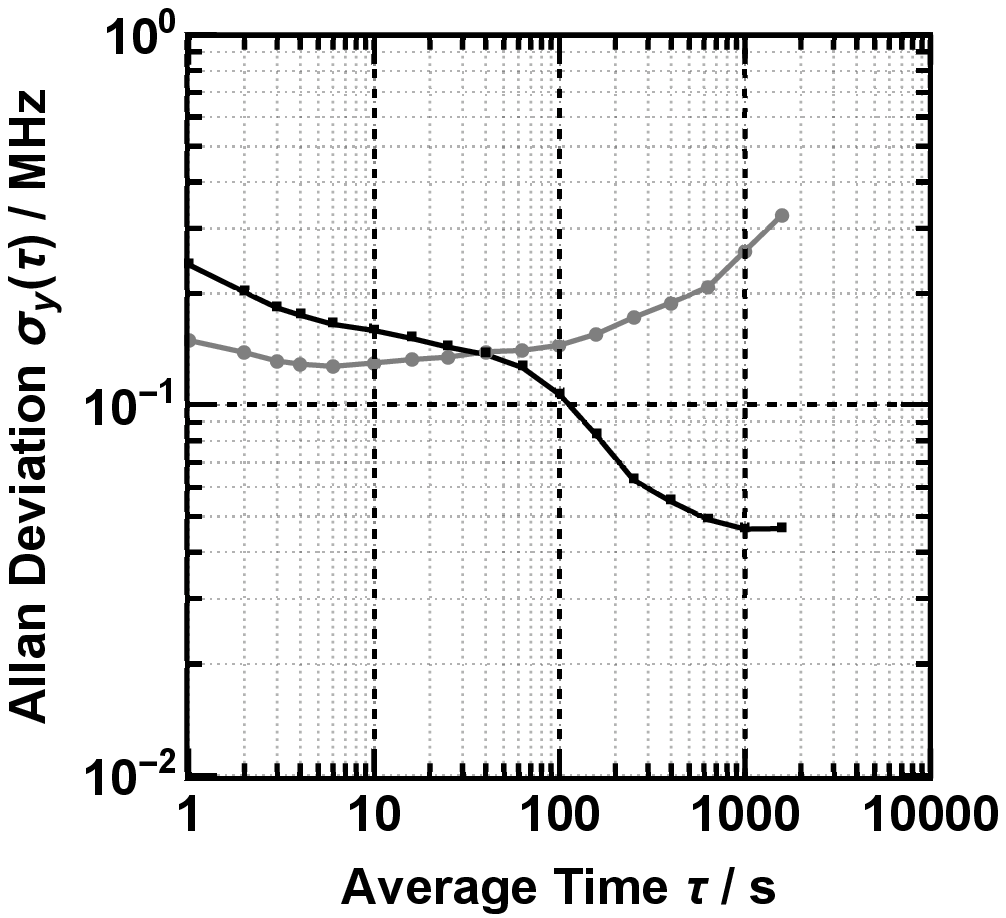}
\caption{Allan deviation $\sigma_y(\tau)$ of the frequency fluctuation either measured by the wavelength meter (gray) or calibrated from the fluorescence counts rates (black) as a function of average time $\tau$.}
\label{fig7}
\end{figure}

The slope of the fluorescence counts rates with respect to detuning frequency around the lock point, $1.2$~KHz/MHz, is also used to convert the fluctuation of the fluorescence counts rates into the 370~nm laser frequency fluctuation after frequency locking. A rough estimation of the frequency stability of the frequency-stabilized 370~nm laser is made by this calibration. The true stability, however, should be obtained from a beat frequency measurement between two identical systems. To characterize the time domain behavior of our frequency-stabilized 370~nm laser, we collect a 3-hours data and calculate the Allan deviation $\sigma_y(\tau)$ of the frequency fluctuation either measured by the wavelength meter or calibrated from the fluorescence counts rates as a function of average time $\tau$, as shown in Fig.~\ref{fig7}. The frequency stability calibrated from the fluorescence counts rates decreases from $239$~KHz at $\tau=1$~s and reaches an almost minimum value of $46$~KHz at $\tau=1000$~s. The frequency stability measured by the wavelength meter first decreases from $149$~KHz at $\tau=1$~s but finally goes up to $261$~KHz at $\tau=1000$~s. This different behavior may indicate a long term drift of the wavelength meter itself. These results show that the frequency-stabilized 370~nm laser described in this paper by using the combined technique is good enough to be employed in the experiments on the laser cooling and trapping of Yb ions.

\section{Conclusion}

In summary, we demonstrate a high-performance frequency stabilization technique for UV diode lasers. Our approach features flexible control, high feedback bandwidth and minimal power consumption of the target UV laser, and is especially suitable for trapped ions quantum experiments, in which the cooling laser is the target UV laser and the ionization laser is the auxiliary laser. The FP cavity can be placed in a vacuum housing to improve atmospheric pressure stability. By using EOM instead of laser diode current for modulation, the linewidth can be further improved. And the DAVLL method may be replaced with the SADL or MTS method with Zeeman level adjustment to obtain a more clearly defined reference frequency.

\bibliography{FS}

\section{Acknowledgements}

This work was supported by the National Natural Science Foundation of China (Grant No.~11704408,~91836106), the Beijing Natural Science Foundation (Grant No.~Z180013) and the Joint fund of the Ministry of Education (6141A020333xx).

The authors thank M.L. Cai and Q.X. Mei for helpful discussions.

\end{document}